\documentclass[pra,showpacs,twocolumn,amsmath,amssymb]{revtex4}

\usepackage{graphicx}%Include figure files
\usepackage{dcolumn}%Align table columns on decimal point
\usepackage{bm}% bold math

\begin{document}

\title{Relativistic coupled-cluster single-double method
applied to alkali-metal atoms}

\author{Rupsi Pal and M. S. Safronova}
\affiliation{Department of Physics and Astronomy, University of
Delaware, Newark, DE 19716-2570, USA}

\author{ W. R. Johnson}
\affiliation{University of Notre Dame, Notre Dame, Indiana 46556,
USA}
\author{Andrei Derevianko}
\affiliation{University of Nevada, Reno, NV 89557-0042, USA}
\author{Sergey G. Porsev}
\affiliation{Petersburg Nuclear Physics Institute, Leningrad district, Gatchina, 188300,
Russia and University of Nevada, Reno, NV 89557-0042, USA}
\begin{abstract}
A relativistic version of the coupled-cluster single-double (CCSD)
method is developed  for atoms with a single valence electron.
In earlier work, a linearized version of the CCSD method (with
extensions to include a dominant class of triple excitations) led to
accurate predictions for energies, transition amplitudes, hyperfine constants, and
other properties of monovalent atoms.  Further progress in high-precision
atomic structure calculations for heavy atoms calls for
improvement of the linearized coupled-cluster methodology.
In the present work, equations for the single and double excitation coefficients
of the Dirac-Fock wave function, including all non-linear coupled-cluster terms
that contribute at the single-double level are worked out.
Contributions of the non-linear terms to energies, electric-dipole matrix elements, and hyperfine
constants of low-lying states in alkali-metal atoms from Li to Cs are evaluated and the
results are compared with other calculations and with precise experiments.

\end{abstract}

\pacs{31.15.Dv %Coupled-cluster theory
      32.10.-f %Properties of atoms
      32.10.Fn %Fine and hyperfine structure
      32.10.Hq %Ionization potentials, electron affinities
      32.70.Cs %Oscillator strengths, lifetimes, transition moments
      }

\maketitle
\section{Introduction}

% defining new commands
\newcommand{\bra}[1]{\langle #1|}
\newcommand{\ket}[1]{|#1\rangle}
\newcommand{\braket}[1]{\langle #1|#1\rangle}

A relativistic version of the coupled-cluster single-double (CCSD) approximation
for monovalent atoms is developed.  In this approximation,
single and double excitations of the (frozen-core) Dirac-Fock wave function for an atom with
one valence electron are included to all orders in perturbation theory.
The relativistic CCSD is an extension of an earlier relativistic single-double (SD) approximation,
in which all nonlinear coupled-cluster terms were omitted. The relativistic SD approximation
provides a method closely related to the configuration-interaction method for
including classes of correlation corrections to Dirac-Fock wave functions
to all-orders in perturbation theory.  When modified to include the dominant triple excitations,
the SD method led to accurate predictions for energies, transition matrix elements,
lifetimes, hyperfine constants, and polarizabilities in alkali-metal atoms
(see, for example, Refs.~\cite{csao,safronova,safrono,safronova1,safrono1,safrono2}).
Owing to recent improvements
in the precision of atomic experiments, it is important to go beyond the relativistic SD approximation and
include the non-linear coupled-cluster terms.

The coupled-cluster method is an all-order extension of many-body perturbation theory
introduced into nuclear physics by Coester and K\"{u}mmel \cite{FC:58,CK:60}
and adapted to atomic and molecular physics by \v{C}i\v{z}ek and Paldus \cite{JC:66,JC:69,CP:71}.
A comprehensive review of the coupled-cluster method and its applications in quantum chemistry
is given in Ref.~\cite{CS:02}.
Relativistic CCSD calculations for monovalent
atoms have been carried out previously
by \citet{AY:94} for transition energies in Li-like U,
by \citet{EE:94e} for ionization and excitation energies of alkali-metal atoms,
by \citet{ISL:99} for polarizabilities of alkali-metal atoms from Li to element 119,
by \citet{AM:00} for charge radii of francium isotopes,
by \citet{LA:01} for electron affinities in alkali-metal atoms Na to element 119,
by \citet{RC:03} for ionization and excitation energies in Rb and Cs,
by \citet{SP:06} for properties of Na, by
\citet{AD:06} for properties of Cs,
and by \citet{BS:06} for parity nonconservation in Cs and Ba$^+$.
Among the nonrelativistic applications of the CCSD method to alkali-metal atoms,
we mention those of \citet{lindgren1} who calculated energies and hyperfine constants of
$2s$ and $2p$ states in Li and  \citet{SY:91}
who evaluated energies and hyperfine constants of $3s$ and $3p$ states in Na.

The relativistic SD method for alkali-metal atoms was introduced in Ref.~\cite{johnson},
where it was used to obtain energy levels, fine-structure intervals, hyperfine constants,
and electric-dipole matrix elements in Li and Be$^+$. Later, the SD method was applied
successfully  to study properties of heavier alkali-metal atoms and monovalent ions
\cite{csao,safronova,safrono,safronova1,safrono1,safrono2}.
Despite the success of the SD all-order method (and its extension to include important triples)
in predicting many properties of monovalent systems,
there are various instances where it fails to produce accurate results. For example,
the magnetic-dipole hyperfine constant for the $7d_{5/2}$ state of Fr
calculated using the SD method
significantly disagrees with the experimental value \cite{grossman}.  The $nd_{5/2}$
hyperfine constants
in Cs provide other such examples \cite{MS-MA10d}. This issue is not limited to the
hyperfine constants; correlation corrections to electric-dipole $6d_j-6p_{j'}$ matrix elements in Rb
are so large that they change the sign of the matrix elements \cite{safronova1}.
The large differences between third-order and all-order
values noted in Ref.~\cite{safronova1} lead to the conclusion that the omitted
higher-order contributions are significant.  Cases where correlation corrections
are extremely large lead to poor accuracy for the SD method, which is generally of high-precision;
in such cases, a more accurate treatment of correlations is mandatory.

Another motivation for further development of the relativistic SD method
is the study of parity nonconservation (PNC) in heavy atoms. One goal of PNC studies is to provide
atomic-physics tests of the standard model of the electroweak interactions through determination of the
weak charge $Q_W$; another is  to extract nuclear anapole moments from PNC measurements.
A precise calculation of the PNC amplitude (together with an uncertainty estimate of the calculation)
is necessary to derive a value of the weak charge $Q_W$ from  experimental measurements.
The accuracy of the  most advanced experimental study of PNC (which was carried out in Cs) is
$0.3\%$~\cite{csPNC}, while the accuracy of the corresponding theoretical calculations is about
0.5\% (see \cite{GF:PNC05} and references therein). The difference between the  value of $Q_W$ extracted from experiment and the
value inferred from the standard model stands at 1$\sigma$ \cite{AD:06}.
More accurate experiments, coupled with improvements in the calculation of
PNC amplitude will lead to a more accurate determination of $Q_W$ and a more stringent
test of the standard model.

Recently, a rather complete treatment of a class of triple excitations that included the valence
electron and two core electrons was carried out
in Refs.~\cite{SP:06,AD:06} for Na and Cs, respectively.
Contributions of quadratic valence non-linear terms were also calculated and found to be relatively large.
As a result, further development of the all-order method must
include a complete treatment of the non-linear terms at the SD level.

In this work, we include all valence and core non-linear coupled-cluster corrections
to the SD equations, leading to a complete set of CCSD equations,  and we study their effects on
various atomic properties of alkali-metal atoms.
In particular,  we calculate ionization
energies and hyperfine constants for the ground $ns$ states and the $np_{1/2}$, $np_{3/2}$ excited states of Li,
Na, K, Rb, and Cs. Reduced electric-dipole matrix elements for the $ns-np_{1/2}$
and $ns-np_{3/2}$ transitions are also calculated.
We give a detailed breakdown of various non-linear contributions in order to identify the
most important effects and  to measure the influence of core non-linear
terms. Comparison of our calculations with other theoretical calculations and with experimental
results is presented.

\section{Coupled-Cluster Method}

In coupled-cluster theory, the wave function of an atom with one valence electron in a state $v$
is written as
\begin{equation}
\ket{\Psi_v} = \exp{(S)}\ket{\Phi_v}, \label{eq1}
\end{equation}
where $\ket{\Phi_v}$ is the lowest-order atomic wave function for atomic state $v$, which
taken to be a frozen core Dirac-Fock (DF) wave function, and where
the wave operator (the operator that maps the DF wave function onto the exact wave function)
is expressed in exponential form  $\exp{(S)}$.
The exponential function in Eq.~(\ref{eq1}) can be expanded to give
\begin{equation}
\ket{\Psi_v}=(1+S+\frac{1}{2}S^2+\cdots)\ket{\Phi_v}.
\label{eqq}
\end{equation}
The cluster operator $S$ is expressed as a sum of $n$-particle excitations $S_n$ of the lowest-order wave function
\begin{equation}
S=S_{1} + S_{2} + \cdots.
\end{equation}
As the number of
excitations increases, the complexity of the wave function increases.
The computational complexity rises dramatically beyond the
 double excitation term $S_2$.  Retaining only single and
double excitations, Eq.~(\ref{eqq}) may be written
\begin{eqnarray}
\nonumber
\ket{\Psi_v}=(1+S_1+S_2+\frac{1}{2}S^2_1+S_1S_2+\frac{1}{6}S^3_1\\
+\frac{1}{2}S^2_2+\frac{1}{2}S^2_1S_2+\frac{1}{24}S^4_1+ \cdots)\ket{\Phi_v}.
\label{SD}
\end{eqnarray}
The one-electron excitation $S_1$ may be either an excitation of a core
electron or an excitation of the valence electron. Correspondingly, the
single core and valence excitations are given by
\begin{eqnarray}
S_{1c} &=& \sum_{ma} \rho_{ma} a^{\dag}_m a_a ,\nonumber\\
S_{1v} &=& \sum_{m\neq v} \rho_{mv} a^{\dag}_m a_v. \label{single}
\end{eqnarray}
Similarly, for double core and valence excitations
\begin{eqnarray}
S_{2c} &=& \frac{1}{2}\sum_{mnab}\rho_{mnab} a^{\dag}_m a^{\dag}_n a_b a_a ,\nonumber\\
S_{2v} &=& \sum_{mnb}\rho_{mnvb} a^{\dag}_m a^{\dag}_n a_b a_v. \label{double}
\end{eqnarray}
The expansion coefficients $\rho_{ma}$ and $\rho_{mv}$  are referred to later as
single core and valence excitation coefficients, while $\rho_{mnab}$ and $\rho_{mnva}$ are
referred to as double core and valence excitation coefficients, respectively.
In Eqs.~(\ref{single}) and (\ref{double}),
 $a^\dagger_i$ and $a_i$ are creation and
annihilation operators for an electron state $i$.
Here and in subsequent formulas, the
indices from the beginning of the alphabet $a,b,...$ designate core states,
indices from the middle of the alphabet $m,n,...$ designate
excited states, the index $v$ labels the valence state, and indices $i,j,k$ and $l$
designate arbitrary states.

In the SD method, only terms linear in the excitation coefficients are retained,
and all remaining terms are omitted.
 Substituting Eqs.~(\ref{single}) and (\ref{double}) into
Eq.~(\ref{SD}) and retaining the terms linear in the excitation
coefficients yields the single-double (SD) all-order wave function
\begin{eqnarray}
 \ket{\Psi_v}=\left[1+ \sum_{ma} \rho_{ma} a^{\dag}_m
a_a +\frac{1}{2}\sum_{mnab}\rho_{mnab} a^{\dag}_m a^{\dag}_n a_b a_a \right. \nonumber\\
\left. +\sum_{m\neq v} \rho_{mv} a^{\dag}_m a_v+\sum_{mnb}\rho_{mnvb}
a^{\dag}_m a^{\dag}_n a_b a_v\right]\ket{\Phi_v} .
\label{SDfunction}
\end{eqnarray}
To derive the equations for  the excitation coefficients (see Ref.~\cite{johnson}
for details),  the SD all-order wave function (\ref{SDfunction}) is substituted into
  the many-body Schr\"{o}dinger equation
  \begin{equation}
H | \Psi_v\rangle=E| \Psi_v\rangle, \label{eq2}
\end{equation}
where the Hamiltonian $H$ is the relativistic {\em
no-pair} Hamiltonian \cite{nopair}, which can be
 written in second-quantized form
as $H=H_0+V$, where
\begin{align}
H_0 & =\ \sum_i \epsilon_i [a_i^\dagger a_i]\, , \label{np1}  \\
V   & =\ \frac{1}{2} \sum_{ijkl}g_{ijkl} [a^\dagger_i a^\dagger_j a_l a_k ] \nonumber \\
    & +\  \sum_{ij} \left(V_\text{DF}  - U\right)_{ij} [a^\dagger_ia_j ]
    + \frac{1}{2} \sum_a (V_\text{DF} - 2 U)_{aa} \label{np2}  \, .
\end{align}
In the {\it no-pair} Hamiltonian,  contributions from
negative-energy (positron) states are omitted. Products of operators enclosed in brackets,
such as $[a^\dagger_ia^\dagger_j a_l a_k ]$, designate normal products with respect to
a closed core.  The quantity $U$ in Eq.~(\ref{np2}) is the model potential used in the
Dirac equation defining single-particle orbitals.
In this work, $U_{ij}$ is taken to be frozen-core Dirac-Fock
potential
\[
U_{ij}=(V_\text{DF})_{ij}=\sum_a(g_{iaja}-g_{iaaj}) .
\]
Such a choice of the potential significantly simplifies calculation
since the second term in Eq.~(\ref{np2}) disappears.
The quantities $g_{ijkl}$ are two-body Coulomb matrix elements:
 \begin{equation*}
g_{ijkl}=\int\!\! d^3r\!\! \int\!\! d^3r^\prime \psi_i^\dagger({\bf r})
 \psi_j^\dagger({\bf r^\prime}) \frac{1}{|{\bf r}-{\bf r^\prime}|}
 \psi_k({\bf r}) \psi_l({\bf r^\prime})
 \end{equation*}
 and the quantity $\epsilon_i$ in Eq.~(\ref{np1}) is the eigenvalue of the Dirac equation.
The third term in (\ref{np2}) is a c-number and provides an additive
constant to the energy of the atom.

The all-order SD wave function given by Eq.~(\ref{SDfunction}) includes only terms that are linear
in the excitation coefficients. In the present work, we take into account all
non-linear terms that arise from the single and double
excitations. Out of all possible non-linear terms, only six terms,
$\frac{1}{2}S^2_1$, $S_1S_2$, $\frac{1}{6}S^3_1$,
$\frac{1}{2}S^2_2$, $\frac{1}{2}S^2_1S_2$, and $\frac{1}{24}S^4_1$
contribute to the single-double equations. Explicitly, the
non-linear terms contributing to the core single-double
equations are $\frac{1}{2}S^2_{1c}$, $S_{1c}S_{2c}$,
$\frac{1}{6}S^3_{1c}$, $\frac{1}{2}S^2_{2c}$,
$\frac{1}{2}S^2_{1c}S_{2c}$, and $\frac{1}{24}S^4_{1c}$ and the
non-linear terms contributing to the valence single-double
equations are $S_{1c}S_{1v}$, $\{S_{1v}S_{2c},S_{1c}S_{2v}\}$,
$\frac{1}{2}S_{1c}^2S_{1v}$, $S_{2c}S_{2v}$, $\{S_{1v}S_{1c}S_{2c}$,
$\frac{1}{2}S_{1c}^2S_{2v}\},$ and $\frac{1}{6}S_{1c}^3S_{1v}$.

First, we consider the contributions from the
non-linear core terms.
The first three non-linear core terms
\begin{eqnarray}
T_1&=&\frac{1}{2}S^2_{1c} = \frac{1}{2}\sum_{rscd}\rho_{rc}\rho_{sd}
a^{\dag}_r a^{\dag}_s a_d a_c , \\
T_2&=&S_{1c}S_{2c} =\frac{1}{2}\sum_{rstcde}\rho_{te} \rho_{rscd}
a^{\dag}_r
a^{\dag}_s a^{\dag}_t a_e a_d a_c ,\\
T_3&=&\frac{1}{6}S^3_{1c} = \frac{1}{6}\sum_{rstcde}\rho_{te}\rho_{rc}
\rho_{sd} a^{\dag}_r a^{\dag}_s a^{\dag}_t a_e a_d a_c
\end{eqnarray}
contribute to equations for both single and double excitation
coefficients,  while the last three terms
\begin{eqnarray}
T_4&=&\frac{1}{2}S^2_{2c} = \frac{1}{8} \sum_{rstucdef}\rho_{rscd}
\rho_{tuef}a^{\dag}_r a^{\dag}_s a^{\dag}_t a^{\dag}_u a_f a_e a_d a_c ,\\
T_5&=&\frac{1}{2}S^2_{1c}S_{2c} \nonumber \\ &=&
\frac{1}{4}\sum_{rstucdef}\rho_{te}\rho_{uf}
\rho_{rscd}a^{\dag}_r a^{\dag}_s a^{\dag}_t a^{\dag}_u a_f a_e a_d a_c ,\\
T_6&=&\frac{1}{24}S^4_{1c} =\frac{1}{24}\sum_{rstucdef}
\rho_{rc}\rho_{sd} \rho_{te}\rho_{uf}a^{\dag}_r a^{\dag}_s
a^{\dag}_t a^{\dag}_u a_f
a_e a_d a_c \nonumber\\
\end{eqnarray}
contribute to the equation for the double excitation coefficients only.

The right-hand side of the single-double equations is obtained by
operating on the non-linear terms above with the two-particle interaction
operator
\begin{equation}
G = \frac{1}{2} \sum_{ijkl}g_{ijkl} [a^\dagger_ia^\dagger_j a_l a_k ]
\end{equation}
according to Eq.~(\ref{np2}).

To derive the equation for the core single-excitation  coefficients, we extract those
terms in $GT_k$ ($k=1,2,3$) that are proportional to $a^\dag_m a_a a^\dag _v$.
To derive the equation for the core double-excitation  coefficients,  we extract the terms in $GT_k$ ($k=1 \cdots 6$)
that are proportional
to $\frac{1}{2}a^{\dag}_m a^{\dag}_n a_b a_a
 a^{\dag}_v$. In all cases, we drop terms corresponding to disconnected
diagrams.
Below, we use the notations $GT^s_k$ and $GT^d_k$  to designate contribution of the corresponding terms to
the single or double excitation equations, respectively.
For clarity, we give the contributions from all terms separately.

The equation for the core single-excitation coefficients becomes
\begin{equation}
(\epsilon_{a}-\epsilon_{m})\rho_{ma}=\text{SD}+GT^s_1+GT^s_2 +GT^s_3,
\label{core_s}
\end{equation}
 where $\epsilon_{i}$ is the one-body DF energy for the state $i$,
  SD is the contribution from the linear
coupled-cluster terms given in \cite{johnson},  and the contributions of the
non-linear terms are
\begin{eqnarray}
GT^s_1&=&
\sum_{drs}\tilde{g}_{mdrs}\rho_{ra}\rho_{sd}-
\sum_{cds}\tilde{g}_{cdas}\rho_{mc}\rho_{sd}, \\
GT^s_2&=&
-\sum_{cdrs}\tilde{g}_{cdsr}\rho_{rsda}\rho_{mc}-
\sum_{cdrs}\tilde{g}_{cdsr}\rho_{smcd}\rho_{ra}\nonumber \\
&+&\sum_{cdrs}\tilde{g}_{cdrs}\tilde{\rho}_{rmca}\rho_{sd},\\
GT^s_3&=&-\sum_{cdrs}\tilde{g}_{cdsr}\rho_{mc}\rho_{rd}\rho_{sa}.
\end{eqnarray}
We used the notation $\tilde{g}_{mnab}=g_{mnab}-g_{mnba}$ and $\tilde{\rho}_{mnab}=\rho_{mnab}-\rho_{mnba}$
in the above formulas.

The equation for the core double-excitation coefficients is
\begin{multline}
(\epsilon_{a}+\epsilon_{b}-\epsilon_{m}-\epsilon_{n})\rho_{mnab}=\text{SD} \\
+GT^d_1+GT^d_2 +GT^d_3+GT^d_4 +GT^d_5+GT^d_6,
\label{core_d}
\end{multline}
where
\begin{eqnarray}
GT^d_1&=& \sum_{rs}g_{mnrs}\rho_{ra}\rho_{sb}+
\sum_{cd}g_{cdab}\rho_{mc}\rho_{nd} \nonumber \\
&-& \left[
\sum_{dr}\tilde{g}_{mdar}\rho_{rb}\rho_{nd} +\left(\begin{array}{c}
   a \leftrightarrow b \\
   m \leftrightarrow n
 \end{array}\right) \right],
\end{eqnarray}
\begin{eqnarray}
GT^d_2&=&
\Biggl[ -\sum_{cdr}\tilde{g}_{cdrb}\rho_{nd}\tilde{\rho}_{rmca}-
\sum_{cdr}\tilde{g}_{cdar}\rho_{rd}\rho_{mncb} \Biggr. \nonumber\\
&+&\sum_{cdr}{g}_{cdra}\rho_{rb}\rho_{nmcd}
+\sum_{crs}\tilde{g}_{ncrs}\rho_{rb}\tilde{\rho}_{smca}\nonumber\\
&+&\sum_{crs}\tilde{g}_{ncrs}\rho_{sc}\rho_{mrab}- \sum_{crs}{g}_{ncrs}\rho_{mc}\rho_{srab}\nonumber \\
&+&\Biggl. \left(\begin{array}{c}
   a \leftrightarrow b \\
   m \leftrightarrow n
 \end{array}\right) \Biggr],
\end{eqnarray}
\begin{eqnarray}
GT^d_3&=&
\Biggl[\sum_{cdr}{g}_{cdar}\rho_{nd}\rho_{mc}\rho_{rb}-
\sum_{crs}{g}_{mcrs}\rho_{nc}\rho_{ra}\rho_{sb}\Biggr. \nonumber\\
&+& \Biggl. \left(\begin{array}{c}
   a \leftrightarrow b \\
   m \leftrightarrow n
 \end{array}\right)\Biggr],
\end{eqnarray}
\begin{eqnarray}
GT^d_4&=&
\sum_{cdtu} g_{cdtu}\rho_{tuab}\rho_{mncd} +\sum_{cdtu}
\tilde{g}_{cdtu}\tilde{\rho}_{mtac}\tilde{\rho}_{undb} \nonumber
\\ &-&\Biggl[\sum_{cdtu} \tilde{g}_{cdtu}(\rho_{tubd}\rho_{mnac}+\rho_{mucd}\rho_{ntba})\Biggr.
 \nonumber\\&+ &
\Biggl.\left(\begin{array}{c}
   a \leftrightarrow b \\
   m \leftrightarrow n
 \end{array}\right)\Biggr],
\end{eqnarray}
\begin{eqnarray}
GT^d_5&=&\sum_{cdtu} g_{cdtu}(\rho_{ta}\rho_{ub}\rho_{mncd}+
\rho_{mc}\rho_{nd}\rho_{tuab}) \nonumber\\
&-&\Biggl[\sum_{cdtu}
\tilde{g}_{cdut}\rho_{tb}\rho_{uc}\rho_{mnad}+ \sum_{cdtu}
\tilde{g}_{cdtu}\rho_{tc}\rho_{nd}\rho_{muab}\Biggr. \nonumber \\
&+& \Biggl. \sum_{cdtu} \tilde{g}_{cdtu}\rho_{tb}\rho_{nc}\tilde{\rho}_{muad})+
\left(\begin{array}{c}
   a \leftrightarrow b \\
   m \leftrightarrow n
 \end{array}\right)\Biggr],
\end{eqnarray}
\begin{equation}
GT^d_6=\sum_{cdtu} g_{cdtu}\rho_{ta}\rho_{ub}\rho_{mc}\rho_{nd}.
\label{final}
\end{equation}
All non-linear contributions to double-excitation coefficients are symmetrized to preserve the
property $\rho_{mnab}=\rho_{nmba}$.

Only one non-linear term, $GT_1=\frac{1}{2}GS^2_{1c}$, contributes to the equation for
the core correlation energy:
\begin{equation}
\delta E_{c}=\delta E_c^{\text{SD}}+\sum_{abmn}\frac{1}{2}\tilde{g}_{abmn}\rho_{ma}\rho_{nb},
\label{core_energy}
\end{equation}
where $\delta E_c^{\text{SD}}$ is the core correlation energy obtained with
linearized SD wave function (\ref{SDfunction}):
\[
\delta E_c^{\text{SD}} = \frac{1}{2} \sum_{mnab} g_{abmn}\tilde{\rho}_{mnab}.
\]
We note that the summation over each index,
for example $i$, involves summing over the principal quantum number
$n_i$, the relativistic angular momentum quantum number
$\kappa_i$, and the magnetic quantum number $m_i$. The sum
over the magnetic quantum numbers is carried out analytically and the
final formulas are given in Appendix \ref{appa}.

The equations for the valence excitation coefficients $\rho_{mv}$ and $\rho_{mnvb}$
are identical to  the core equations given by Eqs.~(\ref{core_s}-\ref{final})
with  replacement of index $a$ by index $v$ and addition of the valence correlation energy $\delta E_v$
into the parenthesis on the left-hand side of both equations, i.e.
\begin{equation}
(\epsilon_{v}-\epsilon_{m}+\delta E_v)\rho_{mv}=\text{SD}+\left( GT^s_1+GT^s_2 +GT^s_3 \right)_{a\rightarrow v},
\label{valence_s}
\end{equation}
\begin{multline}
(\epsilon_{v}+\epsilon_{b}-\epsilon_{m}-\epsilon_{n}+\delta E_v)\rho_{mnvb}=\text{SD}
+ \left( GT^d_1 \right.\hspace{2em}\\
\left. +GT^d_2 +GT^d_3+GT^d_4 +GT^d_5+GT^d_6\right)_{a\rightarrow v}. \label{valence_d}
\end{multline}

The valence correlation energy $\delta E_v$ is given by
\begin{eqnarray}
\delta E_{v}&=& \delta E^{\text{SD}}_{v}
-\sum_{cdt}\tilde{g}_{cdvt}\rho_{td}\rho_{vc}+
\sum_{dtu}\tilde{g}_{vdtu}\rho_{tv}\rho_{ud} \nonumber \\
&-&\sum_{cdtu}\tilde{g}_{cdut}\rho_{vc}\rho_{utvd}-
\sum_{cdtu}\tilde{g}_{cdut}\rho_{tv}
\rho_{uvcd}\nonumber\\&+&\sum_{cdtu}\tilde{g}_{cdtu}\tilde{\rho}_{vtvc}\rho_{ud}
-\sum_{cdtu}\tilde{g}_{cdut}\rho_{td}\rho_{uv}\rho_{vc}.
\label{valence_energy}
\end{eqnarray}
The term $\delta E^{\text{SD}}_{v}$ represents the expression for the valence correlation
energy without the non-linear terms \cite{johnson}:
\[
\delta E^{\text{SD}}_{v} = \sum_{ma} \tilde{g}_{vavm}\rho_{ma} + \sum_{mab} g_{abvm} \tilde{\rho}_{mvab} +
\sum_{mnb} g_{vbmn} \tilde{\rho}_{mnvb}.
\]
The non-linear contributions to the valence correlation energy arise from the
$S_{1c}S_{1v}$, $\{S_{1v}S_{2c},S_{1c}S_{2v}\}$, and $\frac{1}{2}S_{1c}^2S_{1v}$ terms.

We solve the SD equations using a finite basis set. Each orbital
wave function is represented as a linear combination of the
B-splines. We consider a radial grid of 250 points within a sphere
of radius 100 a.u.  We include 35 out of 40 basis orbitals
for each angular momentum and include all partial waves with
$l \leq 6$ in our calculations. A detailed
description of the B-spline method is given in Ref.~\cite{Bspline}.
We treat the non-linear terms on the same footing with the linear
terms, i.e.  all the linear and non-linear terms are
iterated together. First, the equations for the single core and
double core excitation coefficients are iterated until the core correlation
energy given by Eq.~(\ref{core_energy}) converges to  relative accuracy $\epsilon=10^{-5}$.
Then, the valence equations are iterated until the valence correlation energy given by
Eq.~(\ref{valence_energy})
converges to the relative accuracy $\epsilon$.
Atomic properties can be evaluated once the values of the excitation
coefficients are known, as briefly described below.

Matrix elements of a one-body operator
$Z=\sum_{ij}Z_{ij}a_i^{\dag}a_j$ are determined using the formula
\begin{equation}
Z_{wv}=\frac{\bra {\Psi_w} Z
\ket{\Psi_v}}{\sqrt{\braket{\Phi_v}\braket{\Phi_w}}}.
\end{equation}
Substituting the expression for the wave function from Eq.~(\ref{SDfunction})
in the above equation and simplifying, one finds
\begin{equation}
Z_{wv}=\delta_{wv}Z_\text{core}+\frac{Z_\text{val}}{\sqrt{(1+N_v)(1+N_w)}},
\end{equation}
where $Z_\text{core}$, $Z_\text{val}$, $N_v$, and $N_w$ are linear or
quadratic functions of the single and double excitation coefficients written out in
Refs.~\cite{johnson,csao}.

In general, the non-linear terms coming from expanding the
exponent in the CC wave function also contribute to the expressions
to matrix elements. Even at the CCSD truncation level,
one encounters an infinite number of such contributions.
A rigorous  method of partial
summation (dressing) of the resulting series was devised in Ref.~\cite{andrei}.
The method is built
upon expanding a product of cluster amplitudes into a sum of
$n$-body insertions.
Although in the present paper we do not include these
direct non-linear contributions to matrix elements,
calculations~\cite{AD:06} show that dressing may contribute as much as a few 0.1\%
to hyperfine constants in Cs.

\section{Results and Discussion}
\subsection{Energies}

\begin{table}
\caption{Contributions of the non-linear (NL) terms to removal
energies for Li, Na, K, Rb, and Cs. A detailed description of all contributions is given in text.
SD designates the correlation corrections to
the energies calculated using the SD method. All
results are in cm$^{-1}$. \label{tab1}}
\begin{ruledtabular}
\begin{tabular}{lrrr} \multicolumn{1}{l}{Li\rule[-1.2ex]{0ex}{3ex}}&
\multicolumn{1}{c}{$2s_{1/2}$}&
\multicolumn{1}{c}{$2p_{1/2}$}&
\multicolumn{1}{c}{$2p_{3/2}$}\\ \hline
 SD   \rule{0ex}{3ex}                                            & 406.0 & 352.1 & 352.0  \\
 Core NL terms                                    & 0.9   & 0.1   & 0.1    \\
 $S_{2c}S_{2v}$                                   & -5.1  & -4.2  & -4.2   \\
 $S_{1c}S_{1v}$, $\{S_{1v}S_{2c},S_{1c}S_{2v}\}$  & -2.6  & -0.9  & -0.9   \\
 Other valence NL SD terms                           & 0.0   & 0.0   & 0.0    \\
 Total                                            & 399.2 & 347.1 & 347.0  \\[0.2ex]
\hline \multicolumn{1}{l}{Na\rule[-1ex]{0ex}{3.4ex}}&
\multicolumn{1}{c}{$3s_{1/2}$}&
\multicolumn{1}{c}{$3p_{1/2}$}&
\multicolumn{1}{c}{$3p_{3/2}$}\\ \hline
 SD   \rule{0ex}{3ex}                            & 1483.3 & 462.0 & 459.8\\
 Core NL terms                                   & 0.5    & 2.6   & 2.5 \\
 $S_{2c}S_{2v}$                                  & -44.3  & -15.3 & -15.3\\
 $S_{1c}S_{1v}$, $\{S_{1v}S_{2c},S_{1c}S_{2v}\}$ & -23.8  & -10.2 & -10.1 \\
 Other valence NL SD terms                       & 0.0    & 0.0   & 0.0 \\
 Total                                           & 1415.6 & 439.1 & 436.9 \\[0.2ex]
\hline \multicolumn{1}{l}{K\rule[-1ex]{0ex}{3.4ex}}&
\multicolumn{1}{c}{$4s_{1/2}$}&
\multicolumn{1}{c}{$4p_{1/2}$}&
\multicolumn{1}{c}{$4p_{3/2}$}\\ \hline
 SD    \rule{0ex}{3ex}                            & 2869.7 & 1114.3 & 1100.5 \\
 Core NL terms                                    & 26.8   & 11.3   & 11.2   \\
 $S_{2c}S_{2v}$                                   & -142.5 & -59.1  & -58.5  \\
 $S_{1c}S_{1v}$, $\{S_{1v}S_{2c},S_{1c}S_{2v}\}$  & -62.9  & -34.4  & -34.3  \\
 Other valence NL SD terms                        & 0.1    & 0.1    & 0.1    \\
 Total                                            & 2691.2 & 1032.2 & 1019.1 \\[0.2ex]
\hline \multicolumn{1}{l}{Rb\rule[-1ex]{0ex}{3.4ex}}&
\multicolumn{1}{c}{$5s_{1/2}$}&
\multicolumn{1}{c}{$5p_{1/2}$}&
\multicolumn{1}{c}{$5p_{3/2}$}\\ \hline
 SD   \rule{0ex}{3ex}                             & 3423.2 & 1301.1 & 1236.3 \\
 Core NL terms                                    & 31.7   & 13.5   & 13.1   \\
 $S_{2c}S_{2v}$                                   & -185.4 & -76.7  & -73.2  \\
 $S_{1c}S_{1v}$, $\{S_{1v}S_{2c},S_{1c}S_{2v}\}$  & -105.0 & -47.9  & -46.3  \\
 Other valence NL SD terms                        & 0.3    & 0.1    & 0.1    \\
 Total                                            & 3164.8 & 1190.1 & 1130.0 \\[0.2ex]
\hline \multicolumn{1}{l}{Cs\rule[-1ex]{0ex}{3.4ex}}&
\multicolumn{1}{c}{$6s_{1/2}$}&
\multicolumn{1}{c}{$6p_{1/2}$}&
\multicolumn{1}{c}{$6p_{3/2}$}\\ \hline
 SD   \rule{0ex}{3ex}                             & 3881.5 & 1618.7 & 1442.3 \\
 Core NL terms                                    & 44.3   & 18.1   & 16.8   \\
 $S_{2c}S_{2v}$                                   & -224.4 & -107.3 & -96.3  \\
 $S_{1c}S_{1v}$, $\{S_{1v}S_{2c},S_{1c}S_{2v}\}$  & -162.4 & -74.6  & -68.3  \\
 Other valence NL SD terms                        & 0.9    & 0.4    & 0.4    \\
 Total                                            & 3539.9 & 1455.3 & 1294.9 \\
\end{tabular}
\end{ruledtabular}
\end{table}

Table \ref{tab1} shows a detailed breakdown of contributions from non-linear terms
to the removal energies of the alkali-metal atoms Li, Na, K, Rb, and
Cs. To illustrate the relative importance of the various terms, we
conducted five separate calculations for each atom.
Each subsequent calculation includes all terms in the previous
calculation together with additional  terms, the effect of which is being determined.
For clarity, we describe each of the calculations below.
\begin{enumerate}
\item Linearized SD calculation, with all non-linear terms omitted. The
results of this calculation are listed in the rows labeled  ``SD''.
\item All non-linear terms are included in the core equations only, all
non-linear valence terms are omitted. The differences of those values and the
SD results are listed in rows labeled ``Core NL terms''.
\item All non-linear terms are included in the core equations and the quadratic term
$S_{2c}S_{2v}$ is included in the valence equations. The differences with the
calculation (2) give the contributions from the non-linear valence term $S_{2c}S_{2v}$
and are listed in the rows labeled accordingly.
\item All non-linear terms are included in the core equations, and all remaining quadratic
terms are included in the valence equations. The differences between those values and the results of the
calculation (3) give the contributions of the $S_{1c}S_{1v}$ and $\{S_{1v}S_{2c},\, S_{1c}S_{2v}\}$
quadratic valence terms.
 \item Final calculation: all non-linear terms are included in the core and valence equations.
 The results of this calculation are listed in the rows labeled ``Total''.
The differences between those values and the results of the
calculation (4) give the contributions of the cubic and quartic
non-linear terms that are listed in rows labeled ``Other valence NL SD terms''.
\end{enumerate}

 In all of the cases considered here,
the addition of the non-linear terms results in a decrease in the
correlation contributions to the removal energies.
From Table \ref{tab1}, we see that while the
contribution of the core non-linear terms to the removal energies is
negligible compared to contribution of the valence non-linear
terms for the ground states of Na, it becomes significant
(over 10\% of the total NL contribution and about 1\%
of the total correlation energy)
for all states of K, Rb, and Cs considered here. Furthermore, the contribution from core NL terms is  opposite in
sign to that from the valence non-linear contribution.

The $S_{2c}S_{2v}$ term gives the dominant nonlinear contribution, as expected.
 More than half of the contribution from
non-linear terms to the  removal energy arises from this term.
However, contributions from other quadratic terms, $S_{1c}S_{1v}$ and $\{S_{1v}S_{2c},S_{1c}S_{2v}\}$,
are also significant.
Finally, the contributions from terms $\frac{1}{2}S_{1c}^2S_{1v}$,
$\{S_{1v}S_{1c}S_{2c}$, $\frac{1}{2}S_{1c}^2S_{2v}\}$, and
$\frac{1}{6}S_{1c}^3S_{1v}$, i.e.  terms that are cubic or
quartic in the excitation coefficients are negligible.  For Li and Na,
contributions from these terms are essentially zero; for K, Rb, and Cs, the
contributions are less than 0.5\% of the total from all
non-linear terms.
The breakdown of contributions from the valence NL terms is essentially identical for all
states of Na, K, Rb, and Cs under consideration. The total contribution of the NL terms to the
correlation energies of the lowest three states of Na, K, Rb, and Cs is remarkably large;
it is about 1.5\% of the linear SD correlation energy
for Li, 5\% for Na, and 9\% for Cs.

\begin{table}
\caption{Comparison of the all-order removal energies with the theoretical results
obtained by the coupled-cluster method \cite{EE:94e,RC:03} and  experiment.
 All the values are in cm$^{-1}$. \label{tab2}}
\begin{ruledtabular}
\begin{tabular}{lrrr}
\multicolumn{1}{l}{Li \rule[-1.2ex]{0ex}{3.2ex}}&
\multicolumn{1}{c}{$2s_{1/2}$}&
\multicolumn{1}{c}{$2p_{1/2}$}&
\multicolumn{1}{c}{$2p_{3/2}$}\\ \hline
   SD   \rule{0ex}{3ex}     &  43494 &   28586  &  28585  \\
   CCSD                     &  43488 &   28581  &  28580\\
   Expt.{~\cite{nist}}      & 43487 & 28584 & 28583  \\
\hline\multicolumn{1}{l}{Na\rule[-1ex]{0ex}{3.2ex}}&
\multicolumn{1}{c}{$3s_{1/2}$}&
\multicolumn{1}{c}{$3p_{1/2}$}&
\multicolumn{1}{c}{$3p_{3/2}$}\\ \hline
   SD   \rule{0ex}{3ex}      &  41444  &  24495  &  24476 \\
   CCSD                      &  41376  &  24472  &  24453 \\
   SD\cite{safronova}        & 41447 & 24494 & 24477 \\ 
   CCSD\cite{EE:94e}          & 41352   & 24465   & 24447   \\
   Expt.~\cite{nist}         & 41449 & 24493 & 24476 \\
\hline \multicolumn{1}{l}{K\rule[-1ex]{0ex}{3.2ex}}&
\multicolumn{1}{c}{$4s_{1/2}$}&
\multicolumn{1}{c}{$4p_{1/2}$}&
\multicolumn{1}{c}{$4p_{3/2}$}\\ \hline
   SD    \rule{0ex}{3ex}      & 35258  &  22219 &   22066 \\
   CCSD                       & 35080  &  22044 &   21984 \\
   SD\cite{safronova}         & 34962  & 22023 & 21966 \\
   CCSD\cite{EE:94e}           & 35028   & 22016 &   21957       \\
   Expt.~\cite{nist}          & 35010 & 22025 & 21967 \\
\hline \multicolumn{1}{l}{Rb\rule[-1ex]{0ex}{3.2ex}}&
\multicolumn{1}{c}{$5s_{1/2}$}&
\multicolumn{1}{c}{$5p_{1/2}$}&
\multicolumn{1}{c}{$5p_{3/2}$}\\ \hline
   SD     \rule{0ex}{3ex}      & 34021  &  21241  &  20994 \\
   CCSD                        &  33762 &   21130 &   20888 \\
   SD\cite{safronova}         & 33649 & 21111 & 20875 \\
   CCSD\cite{EE:94e}            & 33721   & 21117   & 20877   \\
   CCSD\cite{RC:03}            & 33603   & 21080   & 20831   \\
   Expt.{~\cite{nist}}         & 33691 & 21112 & 20874 \\
\hline \multicolumn{1}{l}{Cs\rule[-1ex]{0ex}{3.2ex}}&
\multicolumn{1}{c}{$6s_{1/2}$}&
\multicolumn{1}{c}{$6p_{1/2}$} &
\multicolumn{1}{c}{$6p_{3/2}$} \\ \hline
   SD    \rule{0ex}{3ex}        & 31871  &  20421 &   19842 \\
   CCSD                         & 31529  &  20258 &   19695 \\
   SD\cite{safronova}           & 31262 & 20204 & 19652 \\
   CCSD\cite{EE:94e}             & 31443   & 20217    & 19669   \\
   CCSD\cite{RC:03}             & 31250   & 20137    & 19574   \\
   Expt.~\cite{nist}            & 31407 & 20228  & 19674 \\
  \end{tabular}
\end{ruledtabular}
\end{table}

In Table~\ref{tab2}, we  compare of our results for the correlation with the linearized all-order 
SD(pT) calculations 
of Ref.~\cite{safronova}, CCSD calculations
of Ref.~\cite{EE:94e},  the CCSD(T) calculations of Ref.~\cite{RC:03}
and with the experimental energies \cite{nist}.
The values in the rows labeled ``SD'' are the sum of lowest-order (DF) energies, the SD
contributions  given in Table \ref{tab1}, and the extrapolated contributions of the higher
partial waves (E$_{\textrm{extrap}}$); the values in the rows labeled
``CCSD'' are the sums of the DF, SD, non-linear contributions,  and E$_{\textrm{extrap}}$.
The CCSD  values should agree with the calculation of Ref.{~\cite{EE:94e}}
within the numerical uncertainties of the calculations.
Significant differences between all-order SD results  and CCSD results
were noted earlier in Ref.~\cite{safronova},
indicating that NL terms may be large. In this work, we find a good agreement
between our CCSD values and the results of Ref.~\cite{EE:94e}.
The remaining discrepancies can be explained by the differences in some
numerical details of the calculations.
In our calculations, we truncated the number of partial waves included in all sums over
excited states at $l_{\text{max}}=6$. Contributions from partial waves with
$l>6$ are extrapolated in second order (see \cite{safrono} for details of the
extrapolation procedure).  The resulting correction increase for heavier alkali-metal atoms
and  is about 0.1\% of the total energy of $6s$ state in Cs.  The differences between
our values and those of Ref.~\cite{EE:94e} are of the same order of magnitude as
the E$_{\textrm{extrap}}$  contributions.

The SD approximation omits contributions to energies of third order in perturbation theory
that arise from triples excitations. These missing third-order terms were included in the 
 calculation of  Ref.~\cite{safronova}. The size of this terms is approximately given by the 
 differences between values listed in rows ``SD'' and ``SD\cite{safronova}''.
  The calculation of Ref.~\cite{EE:94e}
omits these terms entirely. Interestingly, these terms are of nearly the same magnitude as the NL terms
 and of the same sign. As a result, both calculation of Ref.~\cite{safronova},
that omitted NL terms, and of Ref.~\cite{EE:94e}, that omitted missing third-order terms, were in quite
good agreement with experiment. It was shown in \cite{SP:06,AD:06}, that actual iterated 
triple contribution significantly differs from the third-order values. For this reason, and  
 to make a clear comparison with 
previous CCSD results, we omit all triples contributions in this calculation. 
 For Rb and Cs, we also list  values from the CCSD(T) calculations of Ref.~\cite{RC:03} in the rows
labeled ``CCSD\cite{RC:03}'' that include contributions from
triple excitations, but omit the odd-parity channels.  

\begin{table}
\caption{Contributions of the non-linear terms to the
$ns_{1/2}-np_{1/2}$ and $ns_{1/2}-np_{3/2}$ reduced
electric-dipole matrix elements for Li, Na, K,
Rb, and Cs. The final values are  compared with experimental results. All values are
given in atomic units ($ea_0$, where $a_0$ is the Bohr radius). \label{tab3}}
\begin{ruledtabular}
\begin{tabular}{lrr}
\multicolumn{1}{l}{Li  \rule[-1.2ex]{0ex}{3ex}}&
\multicolumn{1}{c}{$2s_{1/2}-2p_{1/2}$}&
\multicolumn{1}{c}{$2s_{1/2}-2p_{3/2}$}\\
\hline
   SD  \rule{0ex}{3ex}                             &  3.31654 & 4.69033 \\
   Core NL terms                                   & -0.00007 &-0.00008 \\
   $S_{2c}S_{2v}$                                  & 0.00063  & 0.00088 \\
   $S_{1c}S_{1v}$, $\{S_{1v}S_{2c},S_{1c}S_{2v}\}$ & 0.00020  & 0.00028 \\
   Other valence NL SD terms                       & 0.00000  & 0.00000 \\
   Total                                           & 3.31730  & 4.69141 \\
   Expt.{~\cite{volz}}                             & 3.317(4) & 4.689(5)\\[0.2ex]
\hline \multicolumn{1}{l}{Na  \rule[-1ex]{0ex}{3.4ex}}&
\multicolumn{1}{c}{$3s_{1/2}-3p_{1/2}$}&
\multicolumn{1}{c}{$3s_{1/2}-3p_{3/2}$}\\
\hline
   SD  \rule{0ex}{3ex}                             & 3.53099 & 4.99314 \\
   Core NL terms                                   & 0.00005 & 0.00006 \\
   $S_{2c}S_{2v}$                                  & 0.00487 & 0.00690 \\
   $S_{1c}S_{1v}$, $\{S_{1v}S_{2c},S_{1c}S_{2v}\}$ & 0.00211 & 0.00297 \\
  Other valence NL SD terms                        & 0.00000 & 0.00000 \\
   Total                                           & 3.53802 & 5.00307 \\
   Expt.{~\cite{volz}}                             & 3.5246(23) & 4.9838(34) \\[0.2ex]
\hline
\multicolumn{1}{l}{K  \rule[-1ex]{0ex}{3.4ex}}&
\multicolumn{1}{c}{$4s_{1/2}-4p_{1/2}$}&
\multicolumn{1}{c}{$4s_{1/2}-4p_{3/2}$}\\
\hline
  SD   \rule{0ex}{3ex}                            & 4.09820  & 5.79392 \\
  Core NL terms                                   &-0.00474  &-0.00669 \\
  $S_{2c}S_{2v}$                                  & 0.02261  & 0.03198 \\
  $S_{1c}S_{1v}$, $\{S_{1v}S_{2c},S_{1c}S_{2v}\}$ & 0.00873  & 0.01231 \\
  Other valence NL SD terms                       &-0.00002  &-0.00003 \\
  Total                                           & 4.12478  & 5.83149 \\
  Expt.{~\cite{volz}}                             & 4.102(5) & 5.800(8)\\[0.2ex]
\hline
\multicolumn{1}{l}{Rb  \rule[-1ex]{0ex}{3.4ex}}&
\multicolumn{1}{c}{$5s_{1/2}-5p_{1/2}$}&
\multicolumn{1}{c}{$5s_{1/2}-5p_{3/2}$}\\
\hline
  SD   \rule{0ex}{3ex}                            & 4.22005 & 5.95527 \\
  Core NL terms                                   &-0.00649 &-0.00913 \\
  $S_{2c}S_{2v}$                                  & 0.03230 & 0.04564 \\
  $S_{1c}S_{1v}$, $\{S_{1v}S_{2c},S_{1c}S_{2v}\}$ & 0.01533 & 0.02156 \\
  Other valence NL SD terms                       &-0.00004 &-0.00006 \\
  Total                                           & 4.26115 & 6.01328 \\
  Expt.{~\cite{volz}}                             & 4.231(3)& 5.977(4)\\[0.2ex]
\hline
\multicolumn{1}{l}{Cs  \rule[-1ex]{0ex}{3.4ex}}&
\multicolumn{1}{c}{$6s_{1/2}-6p_{1/2}$}&
\multicolumn{1}{c}{$6s_{1/2}-6p_{3/2}$}\\
\hline
  SD   \rule{0ex}{3ex}                            & 4.48157 & 6.30391 \\
  Core NL terms                                   &-0.01057 &-0.01482 \\
  $S_{2c}S_{2v}$                                  & 0.04585 & 0.06437 \\
  $S_{1c}S_{1v}$, $\{S_{1v}S_{2c},S_{1c}S_{2v}\}$ & 0.02762 & 0.03865 \\
 Other valence NL SD terms                        &-0.00015 &-0.00021 \\
  Total                                           & 4.54432 & 6.39190 \\
  Expt.{~\cite{rafac}}                            & 4.4890(65) & 6.3238(73)\\
  \end{tabular}
\end{ruledtabular}
\end{table}

\subsection{Reduced electric-dipole matrix elements}

Table~\ref{tab3} gives the
detailed breakdown of the contributions of the non-linear terms to the reduced
electric-dipole matrix elements for Li, Na,
K, Rb, and Cs.  This table is structured in exactly the
same way as Table~\ref{tab1}; the only exception is that we added lowest-order
 DF value to the linearized SD values in rows labeled ``SD''.
 The breakdown of the non-linear terms is identical
to the breakdown for correlation energies. The major contribution comes
from the term $S_{2c}S_{2v}$ as in the case of the removal energies.
Also, there is almost no contribution from terms which are
cubic or quartic in the excitation coefficients. The core
non-linear terms and the cubic and quartic valence non-linear
terms  decrease the E1 reduced matrix elements, while
the quadratic valence non-linear terms increase these
values.
The only exception to the statement above is the contribution of the core
non-linear terms for Na which behaves in the opposite manner.
Since the contribution of the quadratic valence non-linear terms outweigh
the contribution of all the other terms, the ultimate effect is an
increase in values of the reduced electric-dipole matrix
elements. As in the case of the correlation energies, the contribution
of the NL terms is rather large, especially for heavy alkalies where it reaches 1.5\%.
The comparison of our results with the experiment \cite{volz,rafac}
is also given in Table~\ref{tab3}. Addition of
non-linear terms results in a deterioration of the agreement with
experimental values as expected from \cite{SP:06,AD:06} owing to significant
cancellation between the valence NL  and valence triple terms. A very interesting
finding of this work is relatively large contribution of the core non-linear terms for
heavy alkalies. While the core non-linear term is entirely negligible for  Na,
it is found to be 0.2\% for Cs. 

\subsection{Hyperfine constants}

\begin{table}
\caption{Values of the Fermi half-density $c$(fm) parameter used for
magnetization distribution and
gyromagnetic ratio $g_I$ (in units of the nuclear magneton) which
were used to calculate hyperfine constants.\label{tab4}}
\begin{ruledtabular}
\begin{tabular}{lll}
\multicolumn{1}{c}{Atom}& \multicolumn{1}{c}{$c$}&
\multicolumn{1}{c}{$g_I$}\\ \hline
 \\[-0.3cm]
$^7$Li & 1.7995 & 2.17093 \\
$^{23}$Na & 2.8853 & 1.4784 \\
$^{39}$K & 3.6108 & 0.260993 \\
$^{85}$Rb & 4.8708 & 0.54136 \\
$^{133}$Cs & 5.6748 & 0.737886 \\
\end{tabular}
\end{ruledtabular}
\end{table}

We calculate the magnetic-dipole hyperfine constants $A$ for the $ns_{1/2}$ ground
states and the $np_{1/2}$, $np_{3/2}$ excited states of Li, Na,
K, Rb, and Cs. The nuclear magnetization density is described by a Fermi
distribution with half-density radius $c$ and 90\%-10\% falloff thickness $t=2.3$~fm.
Table \ref{tab4} lists values of the
parameter $c$(fm) used for magnetization distribution and the gyromagnetic ratio
$g_I$ for each of the alkali-metal atoms. The nuclear magnetic moments were taken from
Ref.~\cite{ragavan} and weighted averages were considered when more
than one value was present.

Contributions of the various non-linear terms to the hyperfine
constants are given in Table \ref{tab5}. The comparison of our
results with experiment \cite{harper,schlecht,ritter,lyons,wijngaarden,yei,rafac1,tanner}
is also given. 
\begin{table*}
\caption{Contributions of non-linear terms to the magnetic-dipole
hyperfine constants A(MHz) of  Li, Na, K, Rb, and
Cs and comparison with experimental results. The experimental values
are from Ref.~\cite{harper} unless noted otherwise.\label{tab5}}
\begin{ruledtabular}
\begin{tabular}{lrrr}
\multicolumn{1}{l}{Li \rule[-1.2ex]{0ex}{3ex} }&
\multicolumn{1}{c}{$2s_{1/2}$}&
\multicolumn{1}{c}{$2p_{1/2}$}& \multicolumn{1}{c}{$2p_{3/2}$}\\
\hline
   SD \rule{0ex}{3ex}                              & 395.232 & 45.176 &-2.291 \\
   Core NL terms                                   &  -0.025 & -0.005 & 0.006 \\
   $S_{2c}S_{2v}$                                  &  -1.439 & -0.212 & 0.167 \\
   $S_{1c}S_{1v}$, $\{S_{1v}S_{2c},S_{1c}S_{2v}\}$ &  -0.183 & -0.002 &-0.011 \\
   Other valence NL SD terms                       &   0.000 &  0.000 & 0.000 \\
   Total                                           & 393.585 & 44.957 &-2.129 \\
   Expt.                                           & 401.75 $^{(a)}$ & 46.17(35)$^{(b)}$ & -3.07(13)$^{(c)}$ \\[0.2ex]
\hline
\multicolumn{1}{l}{Na \rule[-1ex]{0ex}{3.4ex}}&
\multicolumn{1}{c}{$3s_{1/2}$}&
\multicolumn{1}{c}{$3p_{1/2}$}&
\multicolumn{1}{c}{$3p_{3/2}$}\\
\hline
   SD  \rule{0ex}{3ex}                              & 888.286 & 95.050 & 18.854 \\
   Core NL terms                                    &  -0.663 & -0.242 & -0.057 \\
   $S_{2c}S_{2v}$                                   &  -2.645 & -1.022 & -0.193 \\
   $S_{1c}S_{1v}$, $\{S_{1v}S_{2c},S_{1c}S_{2v}\}$  &  -4.878 & -0.696 & -0.174 \\
  Other valence NL SD terms                         & 0.001   &  0.000 &  0.001 \\
   Total                                            & 880.101 & 93.090 & 18.431 \\
   Expt.                                            & 885.8   & 94.44(13)$^{(d)}$ & 18.534(15)$^{(e)}$\\[0.2ex]
\hline
\multicolumn{1}{l}{K \rule[-1ex]{0ex}{3.4ex}}&
\multicolumn{1}{c}{$4s_{1/2}$}&
\multicolumn{1}{c}{$4p_{1/2}$}&
\multicolumn{1}{c}{$4p_{3/2}$}\\
\hline
  SD  \rule{0ex}{3ex}                               & 237.159  & 28.696 &  6.214 \\
  Core NL terms                                     &   0.457  &  0.138 &  0.032 \\
  $S_{2c}S_{2v}$                                    &  -2.783  & -0.798 & -0.123 \\
  $S_{1c}S_{1v}$, $\{S_{1v}S_{2c},S_{1c}S_{2v}\}$   &  -3.208  & -0.489 & -0.145 \\
 Other valence NL SD terms                          &   0.007  &  0.002 &  0.001 \\
  Total                                             & 231.632  & 27.549 &  5.979 \\
  Expt.                                             & 230.85   & 28.85(30) & 6.09(4)\\[0.2ex]
\hline
\multicolumn{1}{l}{Rb \rule[-1ex]{0ex}{3.4ex}}&
\multicolumn{1}{c}{$5s_{1/2}$}&
\multicolumn{1}{c}{$5s_{1/2}$}&
\multicolumn{1}{c}{$5p_{3/2}$}\\
\hline
  SD  \rule{0ex}{3ex}                               & 1051.554 & 125.624 & 25.560 \\
  Core NL terms                                     &    5.319 &   0.710 &  0.155 \\
  $S_{2c}S_{2v}$                                    &  -14.198 &  -4.091 & -0.577 \\
  $S_{1c}S_{1v}$, $\{S_{1v}S_{2c},S_{1c}S_{2v}\}$   &  -22.636 &  -3.058 & -0.816 \\
 Other valence NL SD terms                          &    0.047 &   0.007 &  0.002 \\
  Total                                             & 1020.086 & 119.192 & 24.324 \\
  Expt.                                             & 1011.9   & 120.7(1)& 25.029(16)\\[0.2ex]
\hline
\multicolumn{1}{l}{Cs \rule[-1ex]{0ex}{3.4ex}}&
\multicolumn{1}{c}{$6s_{1/2}$}&
\multicolumn{1}{c}{$6p_{1/2}$}&
\multicolumn{1}{c}{$6p_{3/2}$}\\
\hline
  SD  \rule{0ex}{3ex}                                & 2439.053 & 311.138 & 51.900 \\
  Core NL terms                                      &   20.455 &   2.604 &  0.475 \\
  $S_{2c}S_{2v}$                                     &  -39.441 & -12.871 & -1.137 \\
  $S_{1c}S_{1v}$, $\{S_{1v}S_{2c},S_{1c}S_{2v}\}$    &  -80.300 & -11.259 & -2.489 \\
 Other valence NL SD terms                           &    0.321 &   0.043 &  0.015 \\
  Total                                              & 2340.088 & 289.655 & 48.764 \\
  Expt.                                              & 2298.2   & 291.89(8)$^{(f)}$ & 50.275(3)$^{(g)}$\\
  \end{tabular}
\end{ruledtabular}
\begin{flushleft}
$^{(a)}$ Reference~\cite{schlecht}
\newline
$^{(b)}$ Reference~\cite{ritter}
\newline
$^{(c)}$ Reference~\cite{lyons}
\newline
$^{(d)}$ Reference~\cite{wijngaarden}
\newline
$^{(e)}$ Reference~\cite{yei}
\newline
$^{(f)}$ Reference~\cite{rafac1}
\newline
$^{(g)}$ Reference~\cite{tanner}
\end{flushleft}
\end{table*}
The structure of  Table~\ref{tab5}
 is identical to that of Table~\ref{tab3}. The ``SD'' values are
 the sum of the lowest-order DF values and correlation correction calculated in the
 absence of the NL terms.  Addition of the non-linear terms
resulted in a decrease in the values of the hyperfine constants in
comparison to the linearized SD values. The NL core terms and the
cubic and quartic valence non-linear terms contribute with a
positive sign to the hyperfine constants, while the quadratic
valence non-linear terms contribute with negative sign for K, Rb,
and Cs. For Na and Li, all core and valence non-linear terms
contribute with the same sign.  The most significant difference
between the NL contributions to the hyperfine constants and to the
removal energies and E1 matrix elements is that the contribution of
the $S_{2c}S_{2v}$ terms is smaller than the contribution of the
other three quadratic valence non-linear terms $S_{1c}S_{1v}$,
$\{S_{1v}S_{2c}$, and $S_{1c}S_{2v}\}$ for the ground states of Na,
K, Rb, and Cs. The breakdown is also different for the $ns$,
$np_{1/2}$, and $np_{3/2}$ hyperfine constants while it is very
uniform for all energies and E1 matrix elements considered here. The
contributions of the core NL terms are particularly large for the
hyperfine constants, almost 1\%(!) of the total values for Cs.

\section{Conclusion}
We have extended the relativistic SD method to include all
non-linear terms at the SD level. The effect of the non-linear terms on the
removal energies, hyperfine constants,  and electric-dipole matrix elements
 of the alkali-metal atoms from Li to Cs was systematically investigated.
In particular, five different calculations were carried out to establish
 the importance of the various contributions for each alkali-metal atom.
  The effect of the core non-linear
  terms was found to be not negligible for heavier alkalies, reaching nearly 1\% of the
  total values of the Cs hyperfine constants. Among other terms, the $S_2^2$
  term was found to be dominant for removal energies and electric-dipole matrix elements with other
  quadratic terms being also significant.
  In the case of the hyperfine constants, the contributions of the other quadratic terms
  exceeded that of the $S^2_2$ term for most cases. The contribution of the cubic and quartic terms
  was found to be negligible in all cases considered in the present work.
 Inclusion of   non-linear terms in the
single-double all-order method is a significant step toward
further development of high-precision methodologies for the calculation
of the atomic properties.

\section{Acknowledgments}
The work of R.P. and M.S.S. was supported in part by National Science Foundation
Grant No. PHY-0457078. The work of W.R.J. was supported in part by National Science Foundation
Grant No. PHY-0456828. The work of A.D. was supported in part by a National
Science Foundation Grant No. PHY-0354876 and by a NIST Precision
Measurement grant. The work of S.G.P. was supported in part
by the Russian Foundation for Basic Research under
Grant No. 05-02-16914-a.

\appendix{
\section{Angular reduction \label{appa}}
\subsection{Designations and definitions}
 The Coulomb interaction $g_{mnab}$ is decomposed into the product of a term $J_k(mnab)$
\begin{multline}
J_k(mmab) = \sum_q (-1)^{j_m-m_m+j_n-m_n+k-q} \times \\
\left(
\begin{array}{ccc}
j_m  & k & j_a\\
-m_m &-q & m_a
\end{array}
\right)
\left(
\begin{array}{ccc}
j_n  & k & j_b\\
-m_n & q & m_b
\end{array}
\right),
\end{multline}
depending on only the angular momentum quantum numbers $j_i$ and $m_i$
of the four states $(m,\, n,\, a,\, b)$,  and a term $X_k(mnab)$ depending only
on the principal quantum numbers $n_i$ and angular quantum numbers $\kappa_i$ of the states:
\begin{equation}
g_{mnab}= \sum_k J_k(mnab)
X_k(mnab).
\end{equation}
Here
\begin{equation}
X_k(mnab)=(-1)^k\left\langle \kappa_m\left\| C^k\right\| \kappa_a\right\rangle \left\langle \kappa _n\left\| C^k\right\| \kappa
_b\right\rangle R_k(mnab). \label{x}
\end{equation}
The quantities $R_k(mnab)$ are (relativistic) Slater integrals and
$\left\langle \kappa _m\left\| C^k\right\| \kappa _a\right\rangle
$ is a reduced matrix elements of a normalized spherical harmonic.
The quantities $Z_k(mnab)$ are given by
\begin{equation*}
Z_k(mnab) = X_k(mnab) + \sum_{k'} [k] \left\{ \begin{array}{ccc}
 j_m  & j_a & k \\
 j_n  & j_b & k'
 \end{array} \right\} X_{k'}(mnba),
\end{equation*}
where $[k]=2k+1$.
Double excitation coefficients have the same angular structure
as Coulomb matrix elements:
\begin{equation*}
\rho_{mnab}= \sum_k J_k(mnab) S_k(mnab)
\end{equation*}
and the quantities $\tilde{S}_k(mnab)$ are defined in the same way
as $Z_k(mnab)$.
The angular
reductions for the single-excitation coefficients are defined as
follows:
\vspace{-1pc}
\begin{eqnarray}
\rho_{ma}=\delta_{\kappa _m\kappa _a}  \delta _{m_m m _a}
     S(ma),
 \nonumber \\ \rho_{mv}=\delta_{\kappa _m\kappa _v}  \delta
_{m_m m _v} S(mv),
\end{eqnarray}
where $\kappa$ is the relativistic angular momentum quantum number
defined as
 $$
\kappa=\mp(j+1/2)\: {\rm for}\: j=l\pm1/2.
$$

\subsection{Angular decomposition of terms contributing to the equation for core
single excitation coefficients}

\begin{eqnarray*}
GT^s_1&=&\sum_{drs}\sqrt{\frac{[j_d]}{[j_a]}}Z_0(mdrs)S(ra)S(sd)\\
&-&\sum_{cds}\sqrt{\frac{[j_d]}{[j_a]}}Z_0(cdas)S(mc)S(sd)\\
GT^s_1&=&-\sum_{cdrsl}\frac{(-1)^{r+s-a-d}}{[l][j_a]}Z_l(cdsr)S_l(rsda)S(mc)\\
&-&\sum_{cdrsl}\frac{(-1)^{c+s-a-d}}{[l][j_a]}Z_l(cdsr)S_l(smcd)S(ra)\\
&+&\sum_{cdrs}\sqrt{\frac{[j_d]}{[j_a]}}\delta_{j_r j_c}Z_0(cdrs)\tilde{S}_0(rmca)S(sd)\\
GT^s_3&=&-\sum_{cdrs}\sqrt{\frac{[j_d]}{[j_a]}}Z_0(cdsr)S(mc)S(rd)S(sa)
\end{eqnarray*}
 \subsection{Angular decomposition of the terms contributing to the equation for core
double excitation coefficients}
 \begin{eqnarray*}
GT^d_1&=&\sum_{rs}X_l(mnrs)S(ra)S(sb)\\
&+&\sum_{cd}X_l(cdab)S(mc)S(nd)\\
&-&\left[\sum_{dr}Z_l(mdar)S(rb)S(nd)+\left(\begin{array}{c}
   a \leftrightarrow b \\
   m \leftrightarrow n
 \end{array}\right)
 \right]
\end{eqnarray*}
 \begin{eqnarray*}
GT^d_2&=&\sum_{cdr}\frac{(-1)^{c+r+l}}{[l]}Z_l(cdrb)\tilde{S}_l(rmca)S(nd)\\
&-&\sum_{cdr}\sqrt{\frac{[j_r]}{[j_a]}}\delta_{j_c j_a}Z_0(cdar)S_l(mncb)S(rd)\\
& +&\sum_{cdrjk}(-1)^{n+m+a+b}[l] \left\{\begin{array}{ccc}
k & l & j\\
a & d & m
\end{array}\right\} \left\{\begin{array}{ccc}
k & l & j\\
b & c & n
\end{array}\right\}\\
& \times&  X_j(cdra)S_k(nmcd)S(rb)\\
&-&\sum_{crs}\frac{(-1)^{l+c+s}}{[l]}Z_l(ncrs)\tilde{S}_l(smca)S(rb)\\
&+&\sum_{crs}\sqrt{\frac{[j_c]}{[j_n]}}\delta_{j_r j_n}Z_0(ncrs)S_l(mrab)S(sc)\\
& -&\sum_{crsjk}(-1)^{n+m+a+b}[l]\left\{\begin{array}{ccc}
k & l & j\\
n & s & b
\end{array}\right\}\left\{\begin{array}{ccc}
k & l & j\\
m & r & a
\end{array}\right\} \\
& \times &X_j(ncrs)S_k(srab)S(mc)+\left(\begin{array}{c}
   a \leftrightarrow b \\
   m \leftrightarrow n
 \end{array}\right)
\end{eqnarray*}

\begin{eqnarray*}
GT^d_3&=& \sum_{cdr}X_l(cdar)S(nd)S(mc)S(rb)\\
&-&\sum_{crs}X_l(mcrs) S(nc)S(ra)S(sb) +\left(\begin{array}{c}
   a \leftrightarrow b \\
   m \leftrightarrow n
 \end{array}\right)
\end{eqnarray*}

\begin{eqnarray*}
GT^d_4&=&\sum_{cdtu}\sum_{k_1k_2k_3k}(-1)^{a+b+t+u}[l][k]\\
&\times&
\left\{\begin{array}{ccc}
k_1 & k_3 & k\\
m & t & c
\end{array}\right\}
\left\{\begin{array}{ccc}
k_1 & k_3 & k\\
n & u & d
\end{array}\right\}
\left\{\begin{array}{ccc}
k_2 & k & l\\
m & a & t
\end{array}\right\} \\
&\times &\left\{\begin{array}{ccc}
k_2 & k & l\\
n & b & u
\end{array}\right\}X_{k_1}(cdtu)S_{k_2}(tuab)S_{k_3}(mncd)\\
&+&\sum_{cdtu}\frac{(-1)^{c+d+t+u}}{[l]^2}Z_l(cdtu)\tilde{S}_l(mtac)\tilde{S}_l(undb)\\
&-&\Bigg[\sum_{cdtuk}\frac{(-1)^{c+d+t+u}}{[l][j_b]}\delta_{j_b j_c}Z_k(cdtu)S_k(tubd)S_l(mnac)\\
&+&\sum_{cdtuk}\frac{(-1)^{c+d+t+u}}{[l][j_m]}\delta_{j_t j_m}Z_k(cdtu)S_k(mucd)S_l(ntba)\\
 &+&\left(\begin{array}{c}
   a \leftrightarrow b \\
   m \leftrightarrow n
 \end{array}\right)
 \Bigg]
\end{eqnarray*}

\begin{eqnarray*}
GT^d_5&=&\sum_{cdtujk}(-1)^{n+m+a+b}[l]\left\{\begin{array}{ccc}
k & l & j\\
b & d & n
\end{array}\right\} \left\{\begin{array}{ccc}
k & l & j\\
a & c & m
\end{array}\right\} \\
&\times & X_j(cdtu) S_k(mncd)S(ta)S(ub)\\
&+&\sum_{cdtujk}(-1)^{n+m+a+b}[l]\left\{\begin{array}{ccc}
k & l & j\\
n & u & b
\end{array}\right\} \left\{\begin{array}{ccc}
k & l & j\\
m & t & a
\end{array}\right\} \\
& \times & X_j(cdtu) S_k(tuab)S(mc)S(nd)\\
&-&\Bigg[\sum_{cdtu}\sqrt{\frac{[j_c]}{[j_b]}}\delta_{j_d j_b}Z_0(cdut)S_l(mnad)S(tb)S(uc)\\
&+&\sum_{cdtu}\sqrt{\frac{[j_c]}{[j_n]}}\delta_{j_n j_u}Z_0(cdtu)S_l(muab)S(tc)S(nd)\\
&+&\sum_{cdtu}\frac{(-1)^{u-d+l}}{[l]}Z_l(cdtu)\tilde{S}_l(muad)S(tb)S(nc)\\
& +&\left(\begin{array}{c}
   a \leftrightarrow b \\
   m \leftrightarrow n
 \end{array}\right)
 \Bigg]
\end{eqnarray*}

\begin{equation*}
GT^d_6=\sum_{cdtu}X_l(cdtu)S(ta)S(ub)S(mc)S(nd)
\end{equation*}

\subsection{Angular decomposition of terms contributing to the core and valence energies}

\begin{equation*}
\delta E^{NL}_c=\frac{1}{2}\sqrt{[j_a][j_b]}Z_0(abmn)S(ma)S(nb)
\end{equation*}

\begin{eqnarray*}
\delta E^{NL}_{v}&=-&\sum_{cdt}\sqrt{\frac{[j_d]}{[j_v]}}Z_0(cdvt)S(td)S(vc)\\
&+&\sum_{dtu}\sqrt{\frac{[j_d]}{[j_v]}}Z_0(vdtu)S(tv)S(ud)\\
&-&\sum_{cdutk}\frac{1}{[k][j_v]}Z_k(cdut)S_k(utvd)S(vc)\\
&-&\sum_{cdutk}\frac{1}{[k][j_v]}Z_k(cdut)S_k(uvcd)S(tv)\\
&+&\sum_{cdtu}\sqrt{\frac{[j_d]}{[j_v]}}Z_0(cdtu)S_0(vtvc)S(ud)\\
&+&\sum_{cdtu}\sqrt{\frac{[j_d]}{[j_v]}}Z_0(cdut)S(td)S(uv)S(vc)
\end{eqnarray*}

%\bibliography{CCSDNL}

\end{document}